\documentclass[aps,prl,twocolumn,superscriptaddress]{revtex4}
\usepackage{amsmath,amssymb}
\usepackage{bm}
\usepackage{graphicx}
\usepackage{epstopdf}

\newcommand{\beq}{\begin{equation}}
\newcommand{\eeq}{\end{equation}}

\begin{document}
\title{Impact of $e-e$ interactions on the superfluid density
	   of dirty
	    superconductors}
\author{V.~A.~Khodel}
\affiliation{National Research Centre Kurchatov
	Institute, Moscow, 123182, Russia}
\affiliation{McDonnell Center for the Space Sciences \&
	Department of Physics, Washington University,
	St.~Louis, MO 63130, USA}
\author{J.~W.~Clark}
\affiliation{McDonnell Center for the Space Sciences \&
	Department of Physics, Washington University,
	St.~Louis, MO 63130, USA}
\affiliation{ Centro de Investiga\c{c}\~{a}o em Matem\'{a}tica
	e Aplica\c{c}\~{o}es, University of Madeira, 9020-105
	Funchal, Madeira, Portugal}
\author{M.~V.~Zverev}
\affiliation{National Research Centre Kurchatov
	Institute, Moscow, 123182, Russia}
\affiliation{Moscow Institute of Physics and Technology,
	Dolgoprudny, Moscow District 141700, Russia}
	
\begin{abstract}
Landau's theory of the Fermi liquid is adapted to analyze the impact of
electron-electron ($e-e$) interactions on the deficit of the superfluid
density $\rho_{s0}=\rho_s(T=0)$ in dirty superconducting electron systems
in which the damping $\gamma$ of single-particle excitations exceeds the
zero temperature BCS gap $\Delta_0$.  In the dirty strong-coupling limit
$\gamma/\Delta_0\gg 1,m^*/m_e\gg 1$, the formula derived for $\rho_{s0}$
is shown to coincide with the well-known  empirical Uemura relation provided
pair-breaking contributions are nonexistent.  The roles of the crystal
lattice and magnetic pair-breaking effects in the observed decline of
the zero-temperature superfluid density $\rho_{s0}$ in overdoped LSCO
compounds are also discussed.  Our method is also applied to elucidation
of results from the pioneering experimental studies performed recently
by Bozovi\`c and collaborators in overdoped LSCO compounds.

\end{abstract}

\maketitle

\section{I.~Introduction}

The phenomenon of high-temperature superconductivity (HTSC), discovered
in two-dimensional electron systems of copper oxides in 1986 \cite{bednorz},
is still a subject of hot debate, defying consistent explanation.  Serious
new challenges are presented by recent experimental studies in overdoped
LSCO compounds~\cite{bozovic,bozltp,bozprl} that reveal an unexpected
deficit of the superfluid density $\rho_{s0}$.

Related discussions~\cite{shag,broun,hirschfeld} of the implications
of this anomalous behavior have focused on the penetration depth
\beq
\lambda_0^2(x)
=(4\pi e^2\rho_{s0}(x)/ m_e)^{-1}
\label{depth}
\eeq
associated with the Meissner effect, which is responsible for the
exponential decay of the external magnetic field at the interior
surface of these compounds at doping values $x$ lower than the critical
value $x_c\simeq 0.3$ at which LSCO superconductivity terminates.

In a major portion of the  phase diagram of the family
La$_{1-x}$Sr$_x$CuO$_4$, but excluding the heavily overdoped region
where $x_c-x\ll x_c$, one is dealing with a type II superconductor,
since the ratio of $ \lambda_0(x)$ to the zero-temperature coherence
length $\xi_0(x)=v_F/\Delta_0(x)$ markedly exceeds unity.  In this
case, the relation between an electric current ${\bf j}$ and the applied
vector potential ${\bf A}$ generating the current, turns out to
be local~\cite{lon,agd,LL9,tinkham}, i.e.,
\beq
j({\bf r})=-\frac{e^2\rho_{s0}}{m_e} A({\bf r})  .
\label{pen}
\eeq

It is a fundamental result of the weak-coupling BCS theory of type II
superconductors that in the clean limit where $\gamma\ll \Delta_0$, the
superfluid density $\rho_{s0}$ coincides with the total electron  density:
\beq
\rho_{s0}=n .
\label{bcslm}
\eeq
Importantly, more sophisticated scrutiny by Larkin and Migdal \cite{lm}
affirms that the relation (\ref{bcslm}) remains unchanged when all
interactions between particles in the normal state are taken into account
within the framework of Landau theory~\cite{lan}.  The same conclusion
was reached later in a different analysis by Leggett~\cite{leggett}.

Conversely, the ratio $\rho_{s0}/n$ in strongly correlated superconducting
electron systems is suppressed, as established in multiple studies beginning
with the well-known article by Uemura et al.~\cite{uemura}.  Among such
studies, the high-quality measurements of Ref.~\cite{bozovic}, performed on
thousands of films of LSCO compounds, are especially valuable, since the loss
of $\rho_{s0}(x)$ has been traced with unprecedented accuracy, warranting
the unambiguous conclusion that the standard BCS approach fails to explain
the new experiments.  It is intriguing that the substantial reduction
of $\rho_{s0}$ persists even at optimal doping $x_o\simeq 0.17$, where
$\rho_{s0}(x_o)\simeq 0.15\,n$~\cite{bozovic}, while upon approach
to the critical value $ x_c\simeq 0.3$, the superfluid density declines
to zero in harmony with the critical temperature $T_c$ -- quite as
if one is dealing with Bose-Einstein condensation (BEC) of bound electron
pairs~\cite{uemura1,shaf,bbook,abook1,abook2}.

However, the observed loss of $\rho_{s0}$ does not require the BEC
phenomenon to be invoked for its explanation.  Rather, in dirty
superconductors where $\gamma>\Delta_0$, such behavior of $\rho_{s0}$
is well documented~\cite{ag,homes,bt}.  It would appear reasonable that
$\gamma(x)$, which grows linearly with doping $x$ due to its
proportionality to the impurity content, may eventually reach values
comparable with the gap $\Delta_0$, especially in the overdoped region
where $T_c(x)\propto\Delta_0(x)$ is known to fall off rapidly with
$x \to x_c$.  It is just such a characteristic dome shape of both
$T_c(x)$ and $\rho_{s0}(x)$ that was uncovered recently in Nb-doped
SrTiO$_3$~\cite{behnia0,thiemann}.

A common objection to applicability of such a dirty-limit scenario to
the LSCO compounds is based on the fact that ARPES data~\cite{shen1}
and numerous observations of Lifshitz-Kosevich oscillations support
a large and well-defined Fermi surface.  However, it follows from
arguments first advanced by L.\ Landau that the presence of a
well-defined Fermi surface and applicability of the dirty limit
are not mutually exclusive~\cite{LL9,LL10}.  In essence, by virtue of
the elasticity of impurity scattering, the problem to be solved
reduces to a quantum-mechanical one of electron motion in an
external potential field, just as in the theory of finite Fermi
systems~\cite{migdal} developed within the framework of Fermi-Liquid
(FL) methods.  Observing that the momentum ${\bf p}$ remains a good
quantum number in crystals, we thus infer that the FL formalism is
applicable to dirty superconductors as well, provided the impact
of damping effects on the structure of the pole part of the electron
Green function is properly taken into account.

Given these conclusions, we now analyze the impact of $e-e$ interactions
as they relate to perplexing behavior exhibited by strongly correlated
electron systems, assuming the onset of superconductivity to be caused
by Cooper pairing with total momentum ${\bf P}=0$. Accordingly, in
calculation of the superfluid density $\rho_s$, we adopt the BCS formalism
without recourse to any alternative propositions.  We concentrate on the
dirty-limit situation $\epsilon^0_F\gg \gamma>\Delta_0$ as described by
the Abrikosov-Gor'kov (AG) theory of superconducting alloys~\cite{ag,agd,LL9}.
As opposed to the clean-limit result (\ref{bcslm}), the superfluid density
is predicted to behave as
\beq
\rho_{s0}(x)\propto n\frac{\Delta_0(x)}{\gamma}.
\label{dl}
\eeq
This implies a corresponding penetration depth in the form
\beq
\lambda_0^2=(4\pi e^2n_s/m_e)^{-1} ,
\label{agpd}
\eeq
with the AG effective density $n_s\propto n\Delta_0/\gamma$ of superfluid
electrons appearing in place of the electron density $n$ in the famous
London formula.

We shall demonstrate that incorporation of the $e-e$ interactions leads to
further loss of superfluid density and  growth of the penetration depth.
This effect has its origin in the presence of a velocity-dependent component
in the amplitude of the effective interaction between quasiparticles, which
is responsible for the enhancement of the effective mass $m^*$ in strongly
correlated electron systems. The result so obtained is shown to be in
agreement with the empirical Uemura relation~\cite{uemura}:
\beq
\lambda_0^2=(4\pi e^2n_s/m^*)^{-1}.
\label{uem}
\eeq
We shall also discuss the pros and cons of the AG pair-breaking scenario,
adopted in Ref.~\cite{broun} to explain the observed change from linear
$\rho_{s0}\propto \Delta_0$ (\ref{dl}) to bilinear $\rho_{s0}\propto
\Delta^2_0$ behavior of the superfluid density upon approach to the
critical doping $x_c$ at which superconductivity terminates.

\section{II.~Generic Formulas for Conventional Fermi Liquids}

We begin by recalling that in BCS theory the electric current
${\bf j}({\bf k})$ is connected with the weak vector potential ${\bf A}$
by
\beq
j_i({\bf k}) =-\frac{ne^2}{m_e}Q_{ij}({\bf k})  A_j({\bf k}) ,
\label{bcsq}
\eeq
where $Q_{ij}({\bf k})=(\delta_{ij}-k_ik_j/k^2)Q(k)$.
Henceforth we adopt the transverse gauge satisfying the condition
$k_j A_j=0$. Thereupon the analysis is simplified considerably, in that
a part of the tensor $Q_{ij}({\bf k})$ emergent from the change of
the gap $\Delta$ in the external magnetic field turns out to be
proportional to the factor $k_ik_j/k^2$ \cite{lm}.  This contribution
to the current ${\bf j}$ then vanishes identically, and we are
left with $Q_{ij}({\bf k})=Q(k)\delta_{ij}$.

The existing weak-coupling BCS-AG theory of superconductivity properly
describes the experimental situation in conventional metals.  However,
this theory fails in strongly correlated electron systems of high-temperature
superconductors. In dealing with the superfluid density, its failure is
evident from comparison of Eqs.~(\ref{agpd}) and (\ref{uemu}) and
is clearly due to the neglect of so-called Fermi liquid (FL) effects
that arise from the fact that the single-particle energy $\epsilon({\bf p})$
is itself a functional of the quasiparticle momentum distribution
$n({\bf p})$.

The magnitude of these FL effects is determined by the variational derivative
$f({\bf p},{\bf p}_1)=\delta\epsilon({\bf p},n({\bf p}_1))/\delta n({\bf p}_1)$,
known in FL theory as the Landau interaction function.  In homogeneous matter,
this phenomenological quantity is identified by a set of parameters, namely
dimensionless harmonics of its Legendre polynomial expansion.  In strongly
correlated Fermi systems, their magnitudes are of order one; hence their
inclusion is imperative.  This can be accomplished in different ways (see,
for example, Ref.~\cite{sr}); however, as we shall see, application of FL
methods to evaluation of the tensor $Q_{ij}$ is advantageous in allowing
us to obtain final results in analytical and persuasive form.

In conventional 3D Fermi liquids where the damping of single-particle
excitations is immaterial, the original FL formula for the tensor $Q_{ij}$
reads (for details, see~\cite{lm}):
\beq
Q_{ij}({\bf k})= \delta_{ij}+\frac{2}{nm_e} \int p_i L({\bf p},{\bf k})
{\cal T} (p_j;{\bf k})\frac {d{\bf p} }{(2\pi)^3}  ,
\label{qk}
\eeq
the particle-hole propagator $L$ being given by the integral
\beq
L({\bf p},{\bf k})=\int[G_s({\bf p}+{\bf k},\varepsilon)
G_s({\bf p},\varepsilon)+F({\bf p}+{\bf k},\varepsilon)
F({\bf p},\varepsilon)]\frac{d\varepsilon}{2\pi i} ,
\label{ipro}
\eeq
where $G_s$ and $F$ are the Gor'kov quasiparticle propagators
\begin{eqnarray}
G_s({\bf p},\varepsilon)&=&\frac{\varepsilon+\epsilon({\bf p})}
{\varepsilon^2-\epsilon^2({\bf p})-\Delta^2({\bf p})},\nonumber\\
F({\bf p},\varepsilon)&=&-\frac{\Delta({\bf p})}{\varepsilon^2
-\epsilon^2({\bf p})-\Delta^2({\bf p})}.
\label{gf}
\end{eqnarray}
For convenience, the factor $z$ identifying the quasiparticle weight
in single-particle states is absorbed into the definition of the
quasiparticle propagators $G_s,F$, and likewise for the vertex part
${\cal T}({\bf p},{\bf k})$, which incorporates FL effects in satisfying
the equation~\cite{agd,LL9}
\beq
{\cal T}({\bf p},{\bf k})={\bf p}+2\int f({\bf n},{\bf n}')
L ({\bf p}',{\bf k}){\cal T}({\bf p}';{\bf k})\frac {d{\bf p}' }{(2\pi)^3},
\label{verfs}
\eeq
where ${\bf n}={\bf p}/p_F$.
For the homogeneous electron liquid, only the first harmonic $f_1$
of the Landau interaction function $f$ enters this equation.

Throughout the whole $T-x$ phase diagram, except for the heavily
overdoped region $|x-x_c|\ll x_c$, the London case $k=0$ applies.
Accordingly, straightforward calculation of the integral
(\ref{ipro}) establishes that the function $L(\epsilon)=L({\bf p};k=0)$
vanishes identically at {\it any momentum} ${\bf p}$ and for {\it any
form of the single-particle spectrum} $\epsilon({\bf p})$~\cite{migdal}.
Such a conclusion remains valid for the tight-binding model spectrum
$\epsilon({\bf p})$ employed in the calculations of $\rho_{s0}$ by
Lee-Hone et al.~\cite{broun}, to guarantee the property
\beq
Q(0)=1 .
\label{lon1}
\eeq
This result immediately triggers {\it recovery} of the BCS-FL relation
(\ref{bcslm}) in superconducting electron systems of solids, provided
damping of single-particle excitations is nonexistent (more details
being provided below).

\section{III.~Fermi-Liquid Effects in Dirty Superconductors}

\subsection{IIIa. Incorporation of $e-e$ interactions}

From the forgoing developments we infer that in the London limit, the
underlying cause of the reduction of $\rho_{s0}$ is the damping of
single-particle excitations in a correlated electron system.  To
facilitate analysis of the impact of FL effects on this reduction,
we first address the case of ordinary impurities where the Anderson
theorem \cite{an} holds, i.e., the critical temperature $T_c$ is not changed
by the presence of impurities.  The textbook dirty-limit formulas,
written for homogeneous matter in the Matsubara representation,
then take the forms~\cite{agd,LL9}
\begin{eqnarray}
G_s(\epsilon,\zeta)& =-&\frac {i\zeta\eta(\zeta)+\epsilon}
{(\zeta^2+\Delta^2_0)\eta^2(\zeta)+\epsilon^2} , \nonumber\\
F( \epsilon,\zeta)& =&\frac{\Delta_0 \eta(\zeta)}{(\zeta^2+\Delta^2_0)
\eta^2(\zeta) +\epsilon^2} ,
\label{grd}
\end{eqnarray}
where
\beq
\eta(\zeta)=1+\frac{\gamma}{2(\zeta^2+\Delta^2_0)^{1/2}} .
\label{eta}
\eeq
It is easily verified that in dirty superconductors, the function
$L(\epsilon,k=0) $, given by Eq.~(\ref{ipro}) with dirty-limit
propagators (\ref{grd}), {\it no longer vanishes}, thus destroying
the coincidence between $\rho_{s0}$ and $n$ that occurs in the
London limit at $\gamma=0$.

In a dirty homogeneous system of interacting electrons, the solution
of Eq.~(\ref{qk}), rewritten in the form
\beq
Q(\gamma)=1+\frac{p_F}{3m_en}{\cal T}_1(p_F,0) L(\gamma)  ,
\label{qgam}
\eeq
is expressed in terms of two quantities:  the particle-hole propagator
$L$ of Eq.~(\ref{ipro}) and the first harmonic ${\cal T}_1$ of vertex
part ${\cal T}$, determined by Eq.~(\ref{verfs}). Their explicit forms
are as follows:
\beq
L(\gamma)= p^2_F \int[G_s(\epsilon,\zeta)G_s(\epsilon,\zeta)
+F(\epsilon,\zeta) F(\epsilon,\zeta)]
\frac{d\zeta d\epsilon d\Omega}{(2\pi)^4v(\epsilon)} ,
\label{lgam}
\eeq
with the group velocity $v(\epsilon)=d\epsilon(p)/dp$ expressed in
terms of the energy $\epsilon$ itself, and
\beq
{\cal T}_1(p_F,0) = p_F[1-f_1L(\gamma)/3]^{-1} ,
\label{tau1}
\eeq
where $f_1$ is the first harmonic of the Landau interaction function.

An inherent problem associated with calculation of $L(\gamma)$ by
Eq.~(\ref{lgam}) is the poor convergence of the integral, an obstacle
usually overcome by subtracting the corresponding result for normal
metals, where $Q(k)$ vanishes~\cite{agd,sr}.  However, this procedure
works flawlessly only in the weak-coupling limit where the vertex
part ${\cal T}$ remains the same in both superconducting and normal
states. Otherwise, an additional contribution proportional to the
corresponding difference of the vertex parts comes into play, introducing
complications.

This obstacle can be surmounted in a different way, aided by the
relation
\beq
\frac{\partial G_s(\epsilon,\zeta)}{\partial\epsilon}=
G_s(\epsilon,\zeta)G_s( \epsilon,\zeta)-F(\epsilon,\zeta)F(\epsilon,\zeta) .
\label{tr}
\eeq
Indeed, upon inserting Eq.~(\ref{tr}) into Eq.~(\ref{lgam}) and performing
some manipulations, we are led to
\beq
L(\gamma)=\frac{2p^2_F}{(2\pi)^3}\left(\int\frac{\partial n(\epsilon)}
{\partial\epsilon}\frac{ d\epsilon d\Omega}{v(\epsilon)}
+ 2 \int \!\!\! \int F^2(\epsilon,\zeta)\frac{d\zeta d\epsilon d\Omega}
{2\pi v(\epsilon)}\right)
\label{lp0}.
\eeq
Both of the integrals involved converge rapidly in the energy
interval of order $\Delta_0$ adjacent to the Fermi surface.
Therefore the group velocity $v(\epsilon)$ can be freely replaced by
the Fermi velocity $v_F$, to arrive finally at
\beq
L(\gamma,\Delta_0)
= -\frac{p_Fm^*}{\pi^2}\Bigl(1-I(\gamma,\Delta_0)\Bigr),
\label{l0}
\eeq
where
\begin{eqnarray}
I(\gamma,\Delta_0)&=&
2 \int \limits_{-\infty}^{\infty} \int\limits_{-\infty}^{\infty}
F^2(\epsilon,\zeta)\frac{d\zeta d\epsilon}{2\pi} \nonumber \\
&=&\Delta^2_0 \int \limits_0^{\infty}\frac{d\zeta}{[\zeta^2+\Delta^2_0]
[(\zeta^2 + \Delta^2_0)^{1/2}+\gamma/2]} .
\label {igamm}
\end{eqnarray}
At small $\Delta_0$, the integrand diverges as $1/\Delta^2_0$, which implies
that the integral $I(\Delta_0)$ varies {\it linearly} with the gap value
as $\Delta_0\to 0$.

It is straightforward to show that accounting for impurity-induced
effects in these equations (as well as those that follow) reduces to the
replacement of the total damping $\gamma$ by its transport version
$\gamma_{\rm tr}$.

With Eqs.~(\ref{l0}) and (\ref{igamm}) in hand, Eq.~(\ref{qgam}) takes
the form
\beq
Q(\gamma,\alpha)=1-\frac{\alpha\Bigl(1-I(\gamma)\Bigr)}
{1+\alpha F_1^0\Bigl(1-I(\gamma)\Bigr)/3}  ,
\label{qint}
\eeq
where $\alpha=m^*/m_e$ and $F_1^0=f_1p_Fm_e/\pi^2$.  Invoking the FL
relation~\cite{LL9,agd}
\beq
m_e/m^*=1-F_1^0/3 ,
\label{effm}
\eeq
the constant $F_1^0$ may be eliminated from Eq.~(\ref{qint}) to yield
\beq
\frac{\rho_{s0}(z,\alpha)}{n}\equiv  Q(z,\alpha)
=\frac {I(z)}{1+(\alpha{-}1) \Bigl(1-I(z)\Bigr)} ,
\label{qfin}
\eeq
where $ z=\gamma/\Delta_0$. The function $I(z)$ can in fact be evaluated
explicitly, with the results \cite{kogan}
\begin{eqnarray}
I(z)&=&\frac {\pi}{z}\left(1+\frac{8\arctan \frac{z-2}{\sqrt{4-z^2}}}
{\pi \sqrt{4-z^2}}\right), \quad z\leq 2, \nonumber\\
I(z)&=&\frac {\pi}{z}\left(1+\frac{8\,{\rm arctanh}
\frac{2-z}{\sqrt{z^2-4}}}{\pi \sqrt{z^2-4}}\right), \quad z>2 .
\label{anal}
\end{eqnarray}

Importantly, Eq.~(\ref{qfin}) simplifies in the dirty strong-coupling
limit $\gamma_{\rm tr}/ \Delta_0\gg 1, \alpha\gg 1$, becoming
\beq
\frac{\rho_{s0}}{n}= \frac{I(z \gg 1)}{\alpha}\simeq
\frac{\pi}{z\alpha}=\pi \frac{\Delta_0}{\gamma_{\rm tr}}
\left(\frac{m_e}{m^*}\right) .
\label{sup1}
\eeq
We conclude that the linear relation between the superfluid density
$\rho_{s0}(x)$ and the gap value $\Delta_0(x)$, emergent in the dirty limit,
comes from the presence of the damping $\gamma$ in the denominator of the
integrand of Eq.~(\ref{igamm}), while incorporation of $e-e$ interactions
leads to a further decline of superfluid density~$\propto m_e/m^*$ relative
to the AG result~\cite{ag}, as documented in Fig.~1.

It is significant that in the strong-coupling dirty limit defined
by $z\gg 1$ and $m^*/m_e\gg 1$, the FL penetration depth, determined
by Eq.~(\ref{depth}), can be rewritten in the Uemura form
\cite{uemura,hashimoto}
\beq
\lambda_0^2=(4\pi e^2n_s/m^*)^{-1}.
\label{uemu}
\eeq
Here $n_s$ stands for the AG superfluid density evaluated with
the aid of Eq.~(\ref{anal}) in the dirty limit $z\gg 1$, while the
additional dependence of $\lambda^2_0$ on the effective mass $m^*$
comes from the $e-e$ interactions.  Thus, in the strong-coupling
dirty limit, the penetration depth $\lambda_0$ {\it diverges}
at the quantum critical point, in tandem with the effective mass.

\begin{figure}[t]
\begin{center}	
\includegraphics[width=0.7\linewidth]{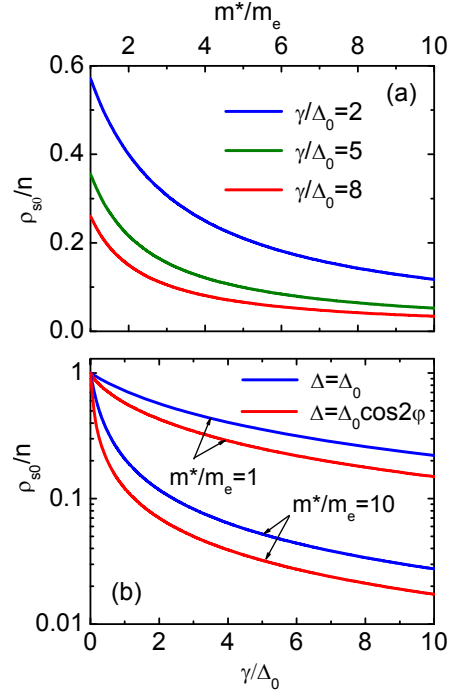}
\end{center}
\caption{Panel (a): Ratio $\rho_{s0}/n$ versus $m^*/m_e$ evaluated from
Eq.~(\ref{qfin}) at $\gamma=2\Delta_0$ (blue line), $\gamma=5\Delta_0$
(green line), and $\gamma=8\Delta_0$ (red line). Panel (b): Ratio
$\rho_{s0}/n$ versus $\gamma/\Delta_0$ evaluated from Eq.~(\ref{qfin})
at $m^*/m_e=1$ and $m^*/m_e=10$ for the $S$-wave gap $\Delta=\Delta_0$
(blue lines) and $D$-wave gap $\Delta=\Delta_0\,\cos2\varphi$ (red lines).}
\label{fig1}
\end{figure}
Let us now turn to the issue of Cooper pairing in copper oxides, where
the gap $\Delta(\phi)$ has $D$-wave structure.  First of all, we observe
that upon keeping the transverse gauge of the vector potential ${\bf A}$
and restricting attention to the first harmonic $f_1$ of the Landau
interaction, the structure of the vertex part ${\cal T}$ itself remains
unchanged, i.e., ${\cal T}( n_i)={\cal T}_1 n_i$. In this situation, the
LM analysis~\cite{lm} informs us that the correction to ${\cal T}_1$
coming from variation of the gap in the external magnetic field is
necessarily proportional to the product $k_i({\bf k}{\bf n})/k^2$ (other
contributions of this variation yielding naught upon multiplication with
the vector ${\bf n}$ and angular integration).  However, the contribution
of this correction to the electric current vanishes identically with
the gauge chosen for the vector potential ${\bf A}$.  In the case of
the D-wave gap, evaluation of Eq.~(\ref{igamm}) with the aid of a
mean-value theorem for integrals yields
`\beq
I_D(\gamma_{\rm av})= \frac{1}{\pi}\int \limits_0^{\infty}\int\limits_0^{\pi}
\frac{\Delta^2_D(\phi) d\zeta d\phi}{[\zeta^2+\Delta^2_D(\phi)]
[(\zeta^2 + \Delta^2_D(\phi))^{1/2}+\gamma_{\rm av}/2]} ,
\label{igamd}
\eeq
where $\gamma_{\rm av}$  stands for an averaged damping value   in the integration interval.

\subsection{IIIb.~Impact of paramagnetic impurities on the relation
between $\rho_{s0}(x)$ and $\Delta_0(x)$}

A comprehensive analysis of this problem, including discussion of
non-Born-limit corrections to the damping $\gamma_s$ (see
Refs.~\cite{rus,hirschfeld2}) and the interplay between Kondo screening
and Cooper pairing (see Ref.~\cite{bauer} and works cited therein) is
beyond the scope of the present work, as its primary aim has been to
demonstrate the importance of $e-e$ interactions in quenching the
superfluid density in copper oxides.  In what follows, we focus on
the role of magnetic effects in the profound change of the dirty-limit
linear relation~(\ref{qfin}) between the superfluid density
$\rho_{s0}(x)$ and the gap value $\Delta_0(x)$, which
prevails over a substantial portion of the LSCO phase diagram, but
yields to bilinear behavior, $\rho_{so}\propto \Delta^2_0$, near the
critical doping $x_c$ at which superconductivity terminates.

A key point underlying this rearrangement is that in overdoped
La$_{1-x}$Sr$_x$CuO$_4$ compounds, the Anderson theorem \cite{an},
which establishes that the gap value is insensitive to the presence of
impurities, and has been substantially involved in the preceding
calculations, no longer holds.  This breakdown is caused by the appearance
of a magnetic component $\gamma_s$ in the damping of single-particle
excitations due to electron scattering on localized Sr magnetic
moments~\cite{ag2,gor}.  Such an effect cannot be absorbed into the
chemical potential, as is done in treating ordinary impurities associated
with the formulas (\ref{grd}).
As a result, each of the denominators of the two Gor'kov propagators $F$
involved in the integrand of Eq.~(\ref{igamm}) acquires an additive
term proportional to the damping $\gamma_s$, which renders them {\it finite}
along with the resulting integral.
Consequently, instead of the linear  relation (\ref{igamm}), a new
behavior applies, namely
\beq
I(\gamma,\Delta_0)\propto\frac{\Delta^2_0}{\gamma_s\gamma_{\rm tr}}.
\label{blr}
\eeq
Accordingly, inclusion of the effects of $e-e$ interactions modifies
the result for $I(\gamma,\Delta_0)$
in the same way as it does for ordinary impurities (cf. Eq.~(\ref{uem})),
implying that
\beq
\frac{\rho_{s0}}{n}\propto\frac{ \Delta^2_0}{\gamma_s\gamma_{\rm tr}\alpha},
\label{blrc}
\eeq
also exploiting the fact that $I\ll 1$ is valid in the relevant region
of the LSCO phase diagram.

\section{IV.~Modification of the FL formalism in crystals}

We next consider how the modified FL approach, adapted above for the
description of damping effects in strongly correlated homogeneous
superconducting electron systems, must be extended to accommodate
lattice-induced phenomena.

\subsection {IVa. Preservation of $\rho_{s0}=n$ equality in the absence
of damping}

It instructive to begin the analysis of crystal-lattice effects with a
validation of the BCS-FL result (\ref{bcslm}) in crystals for the
conventional FL situation in which the damping $\gamma$ vanishes. In
this case, the bare Green function associated with propagation of electrons
in the external field of the crystal lattice has the common form
\beq
G({\bf r}_1,{\bf r}_2, \varepsilon)=\sum \frac{\psi_{\bf p}({\bf r}_1)
\psi_{\bf p}({\bf r}_2)}{\varepsilon- \epsilon({\bf p})
+i\delta\,{\rm sgn}(\varepsilon)}
\label{green}
\eeq
in terms of the corresponding Bloch wave functions $\psi_{\bf p}({\bf r})$.
Analogous expressions apply to the Gor'kov propagators $G_s$ and $F$, with
matrix elements $G_s({\bf p},\varepsilon)$ and $F({\bf p},\varepsilon)$
given by Eq.~(\ref{gf}).  Calculations in which integration is performed
in coordinate space are greatly simplified in the London limit, where only
matrix elements of propagators $G_s,\,F$ evaluated at the same momentum
${\bf p}$ produce a nonzero result, which in fact coincides with
Eq.~(\ref{gf}) by virtue of the orthogonality of different Bloch
wave functions.  Moreover, as before, integration over $\varepsilon$
leads to nullification of the propagator $L$, thus again yielding
$Q_{ij}(0)=\delta_{ij}$.  Accordingly, the BCS-FL result $\rho_{s0}=n$
is recovered in the standard FL case $\gamma = 0$.

\subsection{IVb. FL prescription for the crystal-lattice case}

Explicit treatment of the effects of damping within the crystal-lattice
system may proceed as follows, focusing on 2D electron systems and employing
the formula $d^2p=dp_tdp_n=p_Fd\epsilon d\phi/v(\phi, \epsilon)$, with
$v=|\nabla\epsilon|$.  Since the propagators $G_s,F$ depend just
on $\zeta$ and $\epsilon$, integrations over $\epsilon$ and $\phi$ may
be separated, facilitating calculation. In particular, consider evaluation
of the first contribution
\beq
L^{(1)}(z)=\frac{1}{2}\int \!\!  \int\frac{\partial n(\epsilon)}
{\partial\epsilon} \frac {p_Fd\phi\, d\epsilon} {v(\phi,\epsilon)}
\label{l1z}
\eeq
to the 2D propagator $L(z)$, which has a form analogous to
Eqs.~(\ref{l0})--(\ref{igamm}).  Since the derivative
$\partial n(\epsilon)/\partial\epsilon$ is peaked at the Fermi surface,
the group velocity $v(\phi,\epsilon)$ can be replaced by the Fermi velocity
$v_F(\phi)=v(\phi,\epsilon=0)$ to yield
\beq
L^{(1)}(z)=-N(0)/2 ,
\label{l1s}
\eeq
where
\beq
N(0)= p_F\int \frac {d\phi}{ v_F(\phi)}
\label{dena}
\eeq
is the real 2D density of states. Similarly, one finds
\beq
L(z)=-N(0)(1-I(z))/2 ,
\label{lz}
\eeq
the function $I(z)$ being given by Eq.~(\ref{igamm}).

Further, observing that in the major share of the $T-x$ phase diagram
of the LSCO compounds, the Fermi line has approximately circular
shape~\cite{shen1}, the relation (\ref{tau1}) remains unchanged,
leading after the requisite manipulations to the following result,
\beq
\frac{\rho_{s0}(z,\alpha_c,x)}{n}=\frac{I(z)}
{1+ (\alpha_c(x)-1)\Bigl(1-I(z)\Bigr)}  ,
\label{qinta}
\eeq
in which $\alpha_c(x)=N(0,x)/N_{FL}^0(0)$ with $N_{FL}^0(0)=m_e/\pi$.

The integrand in Eq.~(\ref{dena}) determining the total density of states
$N(0)$ is calculated on the basis of a modified Pitaevskii
equation~\cite{pit,agd}
\beq
{\bf v}({\bf n})={\bf v}_0({\bf n})+2\int f ({\bf n},{\bf n}_1)
\frac {\partial n({\bf p}_1)}{\partial {\bf p}_1}
\frac{d^2{\bf p}_1}{(2\pi)^2}
\label{pit}
\eeq
for the group velocity ${\bf v}({\bf p})=\nabla\epsilon({\bf p})$,
adapted to the 2D case.  The free term ${\bf v}_0({\bf n})$ is the
sum of the gradient of a lattice-induced electric field and
the so-called $\omega$ limit of the vertex part, determined as
${\cal T}({\bf p},\omega\to 0, k\to 0,kv_F/\omega\to 0)$.  In
the homogeneous liquid, where the momentum ${\bf p}$ commutes with
the total Hamiltonian of the problem, this term coincides with
the corresponding Landau result ${\bf p}/m_e$ \cite{pit,migdal}.
In LSCO compounds, ${\bf v}_0({\bf n})$ is replaced by
$\nabla\epsilon_{\rm ARPES}({\bf p})$ \cite{broun}, whose
parameters are extracted from available ARPES data \cite{shen1}.
Evidently, by virtue of the presence of the crystal lattice,
the momentum ${\bf p}$ ceases to commute with the total Hamiltonian,
which, strictly speaking, leads to the occurrence of gradients of
the external potential in the right side of Eq.~(\ref{pit}),
and hence to some renormalization of the term $p/m_e$ appearing
in the corresponding Landau equation.  To avoid further
complications, such contributions are hereafter neglected.

The second term of Eq.~(\ref{pit}), whose integrand contains the Landau
interaction function $f$, accounts for the functional dependence of the
single-particle spectrum $\epsilon({\bf p})$ on the quasiparticle momentum
distribution $n({\bf p})$.  Comparison of Eqs.~(\ref{sup1}) and (\ref{qinta})
demonstrates that in solids, the structure of the Uemura relation remains
unchanged, with the real density of states $N(0)$ absorbing both the
lattice-induced and interaction-induced effects.

There is a widespread belief that elucidation of the electronic properties
of crystals within FL theory is impossible, since its basic equation
relating the single-particle spectrum and the quasiparticle momentum
distribution was derived by Landau under the assumption of Galilean
invariance, which breaks down for electrons inhabiting crystals.  We
observe, however, that Eq.~(\ref{pit}) is almost identical to the original
Landau equation~\cite{agd}. The crucial distinction is that this
equation has instead been derived on the basis of gauge invariance,
which is known to hold in crystals as well as homogeneous systems.
Accordingly, it is the Pitaevskii equation that should be employed
in calculations of the real density of states $N(0)$, which is
responsible for renormalization of the AG results in interacting
electron systems of solids.

\section{VI.~Discussion and Summary}

Let us first consider the situation that prevails in the London limit
for conventional Fermi liquids in which the damping $\gamma$ of
single-particle excitations is negligible.  In this case, it follows
from our analysis that the superfluid density $\rho_{s0}$ must coincide
with the total electron density $n$, irrespective of the form of the
single-particle spectrum, which is naturally quite complicated due
to lattice effects and the character of the interactions between
particles, sometimes giving rise to the occurrence of flat portions
in $\epsilon({\bf p})$.

In the case of dirty homogeneous superconductors with the conventional
FL ground state, introduction of $e-e$ interactions has been shown
to alter the elementary AG behavior, in which the loss of superfluid
density $\rho_{s0}$ depends solely on the AG parameter
$z=\gamma/\Delta_0$~\cite{ag,kogan}.  As illustrated in panel (a)
of Fig.~1, an additional decline of $\rho_{s0}$ is found to be
triggered by the presence of effective velocity-dependent interactions
between quasiparticles that produce an enhancement of the density of
states $N(0)$ associated with the ratio $m^*/m_e$.  We have demonstrated
that the Uemura relation $\lambda_0^2=(4\pi e^2n_s/m^*)^{-1}$ does in
fact apply in the strong-coupling dirty limit $z\gg 1$, $m^*/m_e\gg 1$,
with (i) the effective mass $m^*$ characterizing the interaction-induced
contribution to $\rho_{s0}$ and (ii) the Uemura parameter $n_s$
representing the AG superfluid density associated with the function
$I(z)$ given by Eq.~(\ref{anal}).

In connection with these results, it is worth noting that in their first
article \cite{broun} devoted to evaluation of the superfluid density of
overdoped LSCO compounds, the authors have sought to explain the unorthodox
behavior of the superfluid density of LSCO compounds uncovered by
Bozovi\`c et al.~\cite{bozovic} within the so-called semiclassical
scheme~\cite{am,chandr}.  In this procedure, the magnetic field is
incorporated by making the replacement ${\bf p}\to {\bf p}-e{\bf A}$
solely in the assumed {\it single-particle spectrum}, whose parameters are
determined from the available ARPES data.  In so doing, FL effects associated
with the variation of the single-particle energy $\epsilon(p)$ due to
the change of the quasiparticle momentum distribution $n(p)$ -- which
are naturally incorporated for the homogeneous electron liquid in Secs.~II
and III above --  are completely ignored within the framework of the
semiclassical scheme.  Without taking proper account of these effects,
whose magnitude increases with growth of the ratio $m^*/m_e$, elucidation
of the Uemura relation~(\ref{uem}) becomes impossible, because the decisive
factor $m^*/m_e$ is lost.

Clearly, this deficiency of the semiclassical approach persists in dealing
with strongly correlated electron systems of high-$T_c$ superconductors
moving in the external field of their crystal lattice.  (As seen from
Eq.~(\ref{pit}), it persists irrespective of whether proper treatment
has been given to the logarithmic-like divergence of the tight-binding
density of LSCO states~\cite{bansil,shen1}, which is exhibited in
the doping region where the Fermi line touches the zone boundary.)
Furthermore, as seen from Eq.~(\ref{pit}), the effects of the lattice
and interaction, working in tandem, change the group velocity profoundly,
and hence the density of states $N(0)$ itself. Consequently, the only
way to proceed without extensive and problematic numerical calculations
based on Eq.~(\ref{pit}) is phenomenological.  Within the modified
FL theory presented here, these effects are naturally absorbed into
a single density-of-states parameter $\alpha$, and the same is true
for the Sommerfeld coefficient in the specific heat $C(T)$.  We
refer to the available experimental information on the LSCO
compounds~\cite{tai1,tai2,horio} to extract the parameter $\alpha$.

In comparing our results with experimental data on the LSCO superfluid
density $\rho_{s0}(x)$, we focus on the overdoped region $0.20<x< 0.25$,
which is free of pseudogap influence~\cite{shen1} and where experimental
values of the key input parameter $\gamma(x)$ are available~\cite{bozprl}.
As is known, the experimental curve $T_c(\rho_{s0})$ consists of a
dominant linear portion and, in a relatively small region adjacent
to the origin, $T_c$ behaves as $\sqrt{\rho_{s0}}$~\cite{bozovic,bozltp}.
The linear segment of the curve is associated with the Uemura-like portion
given by the theory as expressed in Eq.~(\ref{sup1} and characterized by its
slope $dT_c/d\rho_{s0}=(2.5\pm 0.1)\times 10^2\,{\rm K}$~\cite{bozovic,bozltp}.
In accord with Eq.~(\ref{sup1}), the slope depends on the product of the
damping $\gamma$ and the density-of-states factor $\alpha$. Since the doping
region involved is quite narrow, one might expect that the $x$-variations
of the two input parameters
involved can be neglected. If so, at the midpoint $x=0.22$ of the doping
interval implicated, where $\gamma=75\, {\rm K}$~\cite{bozprl} is known
from experimental data, simple numerical calculations based on
Eq.~(\ref{igamd}) derived for $D$-pairing, as appropriate the LSCO
compounds, yield a theoretical slope of $ 2.2\times 10^2\,{\rm K}$.  This
is close to the experimental value $(2.5\pm 0.1)\times 10^2\,{\rm K}$,
provided the BCS relation $2\Delta_0=4.28T_c$ is adopted and the effective
mass value is chosen to be $m^*=12m_e$, in accordance with the relevant
experimental data~\cite{tai2}.

However, the issues raised by the hypothetical assumption of permanent
input parameters as functions of doping are more involved. In the doping
range under consideration, the damping $\gamma(x)$ is doubled, increasing
linearly toward $ x_c$~\cite{bozprl}.  Moreover, the change of $\alpha(x)$
associated with the aforementioned logarithmic divergence of the
tight-binding density of states on the left edge on the doping interval,
occurring at a critical value $x_t\simeq 0.2$~\cite{shen1}, is even more
profound.  Fortunately, the variations of $\gamma(x)$ and $\alpha(x)$
swing in opposite directions, thereby suppressing the net change of
$\rho_{s0}(x)$ and allowing it to be neglected in a first approximation.
The next step toward improving the reliability of the results obtained
within the extended FL approach would involve numerical solution of
Eq.~(\ref{pit}) to obtain a realistic quasiparticle group velocity
$v({\bf p})$ for insertion into the integral (\ref{igamm}). The energy
dependence of the damping $\gamma$, known from experiment~\cite{bozovic},
should be properly taken into account as well.

In explanation of the second segment of $\rho_{s0}(x)$, characterized
by its bilinear dependence on $\Delta_0$ and situated adjacent to
the critical doping $x_c$, an idea advanced many years ago by
Abrikosov and Gor'kov \cite{ag2} has been invoked to attribute the
rearrangement of the linear regime to the presence of a {\it magnetic}
part $\gamma_s$ of the damping of single-particle excitations.
In their original model, the authors of Ref.~\cite{ag2} considered a
pair-breaking mechanism associated with electron scattering by impurity
magnetic moments.  Within this model, a particular behavior
$T^2_c(x)\propto x_c-x$ observed experimentally is reproduced.

However, the application of this idea to elucidation of the available LSCO
experimental data~\cite{bozovic,bozprl,bozltp} encounters some difficulties.
For example, these experiments have shown no trace of gapless
superconductivity, which is an integral feature of the AG pair-breaking
mechanism. Moreover, the BCS approach fails to explain basic features of
high-temperature superconductivity, including the enhancement of
the critical temperature $T_c$ itself.  In this situation, results
from application of the BCS gap equation to the problem appear to
be inconsistent.  Given these considerations, the version of the AG
paramagnetic scenario adopted in Ref.~\cite{broun} becomes questionable.

A potential source of the observed discrepancy of predictions of
extended FL theory applied here from the experimentally established
behavior of the superfluid density $\rho_{s0}\propto T^2_c$
upon approach to critical doping, as well as the
challenging temperature dependence of the superfluid density
$\rho_s(T)$~\cite{bozovic,bozltp}, may be related to a rearrangement of
normal states of strongly correlated electron systems associated with
violation of their topological stability~\cite{ks}.  Such a phenomenon
is now actively discussed following publication of a series of articles
devoted to the occurrence of {\it flat bands} in magic-angle twisted
bilayer graphene~\cite{jh1,jh2,jh3,yankovitz}. In the same vein, we
may point to the recent observation \cite{arnold} of a magnetic-field
dependent electronic gap in the point-contact spectrum of dirty
graphite.  This observation is indicative of local superconductivity
having an estimated critical temperature $T_c \approx 14~{\rm K}$,
with possible implication of a flat-band mechanism \cite{volovik1,volovik2}.
Remarkably, it is in exactly the present case of {\it overdoped LSCO
compounds} considered here that arguments favoring the emergence and
agency of flat bands in strongly correlated electron systems of cuprates
have recently been reiterated in Ref.~\cite{PLA2018}. In future work,
we plan to investigate the role that flat bands may have in quenching
the superfluid density $\rho_s(T)$ and in its unexpected temperature
dependence.

To summarize, we have demonstrated that the basic regime of behavior of
the LSCO superfluid density $\rho_{s0}(T_c)$, where it changes linearly
with $T_c$, is properly reproduced within the AG-FL theory, the calculated
slope being in agreement with experiment.  As for the second regime,
operative near critical doping $x_c$ where $\rho_{s0}(T_c)\propto T^2_c$,
effort toward its quantitative explanation remains inconclusive.

\section{Acknowledgments}
We thank I. Bozovi\`c, V.\ Kogan, Ya.\ Kopelevich, and G.\ Volovik for
fruitful discussions and V.\ Stephanovich for valuable comments. VAK and
JWC acknowledge financial support from the McDonnell Center for the
Space Sciences.


\begin{thebibliography}{99}

\bibitem{bednorz} J.\ G.\ Bednorz, K.\ A.\ M\"uller,  Z.\ Phys.\ B {\bf 64},
189 (1986).	
	
\bibitem{bozovic} I.\  Bozovi\`c, X.\ He, J.\ Wu., A.\ T.\ Bollinger,
Nature {\bf 536}, 309 (2016).

\bibitem{bozltp} I.\ Bozovi\`c, A.\ T.\ Bollinger, J.\ Wu, X.\ He, Low.\ Temp.\ Phys.\ {\bf 44}, 519 (2018).

\bibitem{bozprl}  F.\ Mahmood, X.\ He, I.\ Bozovi\`c, N.\ P.\ Armitage, Phys. Rev. Lett. {\bf 122}, 027003 (2019).

\bibitem{shag} V.\ R.\ Shaginyan, V. A. Stephanovich, A.\ Z.\ Msezane,
G.\ S.\ Japaridze, K.\ G.\ Popov et al., Phys. Chem. Chem. Phys. {\bf 19},
21964 (2017).

\bibitem{broun} N.\ R.\ Lee-Hone, J.\ S.\ Dodge, D.\ M.\ Broun, Phys. Rev. B {\bf 96} 024501 (2017).

\bibitem{hirschfeld} N.\ R.\ Lee-Hone, V.\ Mishra, D.\ M.\ Broun, P.\ J.\ Hirschfeld, arXiv:1802.10198.



\bibitem{lon} F.\ and H.\ London, Proc.\ Roy.\ Soc.\ (London) A {\bf 149},
71 (1935).

\bibitem{agd} A.\ A.\ Abrikosov, L.\ P.\ Gor'kov, I.\ E.\ Dzjaloshinski,
{\it Methods of Quantum Field Theory in Statistical Physics} (Pergamon
Press, Oxford, 1965).

\bibitem{LL9} L.\ D.\ Landau, E.\ M.\ Lifshitz, {\it Statistical Physics},
Vol.\ II (Pergamon Press, Oxford, 1980).

\bibitem{tinkham} M.\ Tinkham, {\it Introduction to Superconductivity}.
(Dover Publications, Mineola NY, 2004).

\bibitem{lm} A.\ I.\ Larkin, A.\ B.\ Migdal, Sov.\ Phys.\ JETP {\bf 17},
1146 (1963).

\bibitem{lan} L.\ D.\ Landau, Sov.\ Phys.\ JETP {\bf 3}, 920 (1957);
{\bf 8}, 70 (1958).

\bibitem{leggett} A.\ J.\ Leggett, J. Stat.\ Phys.\ {\bf 93}, 927 (1998).

\bibitem{uemura} Y.\ J.\ Uemura et al., Phys.\ Rev.\ Lett.\ {\bf 62},
2317 (1989).

\bibitem{uemura1}  Y.\ I.\ Uemura, Physica C {\bf 282-287} Part 1, 194 (1997).

\bibitem{shaf} M.\ R.\ Shafroth, Phys.\ Rev.\ {\bf 96}, 1442 (1954).

\bibitem{bbook}  J.\ M.\ Blatt, {\it Theory of Superconductivity}, (Academic
Press, New York, 1964).

\bibitem{abook1} A.\ S.\ Alexandrov, N.\ F.\ Mott, {\it Polarons and
Bipolarons} (World Scientific, Singapore, 1996).

\bibitem{abook2} A.\ S.\ Alexandrov, {\it Theory of Superconductivity:
From Weak to Strong Coupling} (IoP Publications, Bristol, 2003).

\bibitem{ag} A.\ A.\ Abrikosov, L.\ P.\ Gor'kov, Sov. Phys. JETP, {\bf 8},
1090 (1959).

\bibitem{homes}
C.\ C.\ Homes, S.\ V.\ Dordevic, M.\ Strongin, D.\ A.\ Bonn, R.\ Liang,
W.\ N.\ Hardy, S. Koymia, Y.\ Ando, G.\ Yu, X.\ Zhao, M.\ Greven,
D.\ N.\ Basov, T.\ Timusk, Nature {\bf 430}, 539 (2004).

\bibitem{bt} D.\ N.\ Basov, T.\ Timusk, Rev.\ Mod.\ Phys.\ {\bf 77},
722 (2005).

\bibitem{behnia0}
C.\  Collignon, B.\ Fauqu'e, A.\ Cavanna, U.\ Gennser, D.\ Mailly,
K. Behnia, Phys.\ Rev.\ B {\bf 96}, 224506 (2017).

\bibitem{thiemann}
M.\ Thiemann, M.\ H.\ Beutel, M.\ Dressel, N.\ R.\ Lee-Hone, D.\ M.\ Broun,
E.\ Fillis-Tsirakis, H.\ Boschker, J.\ Mannhart, M.\ Scheffler, Phys.\
Rev.\ Lett.\ {\bf 120}, 237002 (2018).

\bibitem{shen1}
T.\ Yoshida, X.\ J.\ Zhou, D.\ H.\ Lu, S.\ Komiya, Y.\ Ando, H.\ Eisaki,
T.\ Kakeshita, S.\ Uchida, Z.\ Hussain, Z.-X.\ Shen, and A.\ Fujimori,
J.\ Phys.\ Condens.\ Matter {\bf 19}, 125209 (2007).

\bibitem{LL10} E.\ L.\ Lifshitz, L.\ P.\ Pitaevskii, {\it Physical Kinetics}
(Pergamon Press, Oxford, 1981).

\bibitem{migdal} A.\ B.\ Migdal, {\it Theory of Finite Fermi Systems and
Applications to Atomic Nuclei} (Wiley, New York, 1967).

\bibitem{sr} J.\ W.\ Serene, D.\ Rainer, Phys. Rep. {\bf 101}, 221 (1983).

\bibitem{an} P.\ W.\ Anderson, J. Chem. Phys. Solids, {\bf 11} 26 (1959).

\bibitem{kogan} V.\ G.\ Kogan, Phys. Rev. B {\bf 87}, 220507(R) (2013).

\bibitem{hashimoto}
K.\ Hashimoto, K.\ Cho, T.\ Shibauchi, S.\ Kasahara, Y.\ Mizukami,
R.\ Katsumata1, Y.\ Tsuruhara, T.\ Terashima, H.\ Ikeda, M.\ A.\ Tanatar,
H.\ Kitano, N.\ Salovich, R.\ W.\ Giannetta, P.\ Walmsley, A.\ Carrington,
R.\ Prozorov, Y.\ Matsuda1, Science {\bf 336}, 1554 (2012).

\bibitem{ag2} A.\ A.\ Abrikosov, L.\ P.\ Gor'kov, Sov. Phys. JETP, {\bf 12},
1243 (1960).

\bibitem{gor} L.\ P.\ Gor'kov,  {\it Theory of Superconducting Alloys}; Superconductivity vol. I,  Ed. K.\ H.\ Bennemann, J.\ B.\ Ketterson, Springer-Verlag 2008.

\bibitem{rus} A.\ I.\ Rusinov, Sov. Phys. JETP, {\bf 29},
1101 (1969).

\bibitem{hirschfeld2} P.\ J.\ Hirschfeld, N.\ Goldenfeld, Phys. Rev. B {\bf 48}, 4219 (1993).

\bibitem{bauer} J.\ Bauer, J.\ T.\ Pascual, K.\ J.\ Franke, Phys. Rev. B {\bf 87} 075125 (2013).



\bibitem{kogan2}  V.\ G.\ Kogan, R.\ Prozorov, V.\ Mishra, Phys. Rev. B
{\bf 88}, 224508 (2013).

\bibitem{am} N.\ W.\ Ashcroft, N.\ D.\ Mermin, {\it Solid State Physics},
(Holt, Rinehart, and Wilson, New York, 1976), Chapter 12.

\bibitem{chandr} B.\ S.\ Chandrasekhar, D.\ Einzel, Annalen der Physik,
{\bf 503}, 535 (1993).

\bibitem{pit} L.\ P.\ Pitaevskii, Sov. JETP {\bf 10}, 1267 (1960).

\bibitem{bansil} S.\ Markiewicz, S.\ Sahrakorpi,  M.\ Lindroos,
H.\ Lin, A.\ Bansil, Phys. Rev. B{\bf 72} 054519 (2005).

\bibitem{tai1} B.\ Michon et al., arxiv:1804.08502

\bibitem{tai2}
A.\ Legros, S.\ Benhabib, W.\ Tabis, F.\ Lalibert\'e, M.\ Dion,
M.\ Lizaire, B.\ Vignolle, D.\ Vignolles, H.\ Raffy, Z.\ Z.\ Li,
P.\ Auban-Senzier, N.\ Doiron-Leyraud, P.\ Fournier, D.\ Colson,
L.\ Taillefer, C.\ Proust, Nat. Phys. doi/org/10.1038; arXiv:1805.02512.

\bibitem{horio} M.\ Horio et al., arxiv:1804.08019v2.

\bibitem{bonn} D.\ M.\ Broun, W.\ A.\ Hutterma, P.\ I.\ Turner,
S.\ Ozcan, B.\ Morgan, R.\ Liang, W.\ N.\ Hardy, D.\ A.\ Bonn,
Phys. Rev. Lett. {\bf 99}, 237003 (2007).

\bibitem{ks} V.\ A.\ Khodel and V.\ R.\ Shaginyan, JETP Lett.\ {\bf 51},
553 (1990).

\bibitem{jh1}
Y.\ Cao, V.\ Fatemi, S.\ Fang, K.\ Watanabe, T.\ Taniguchi, E.\ Kaxiras,
P.\ Jarillo-Herrero, Nature {\bf 556}, 43 (2018).

\bibitem{jh2}
Y.\ Cao, V.\ Fatemi, A.\ Demir, S.\ Fang, S.\ L.\ Tomarken, J.\ Y.\ Luo,
J.\ D.\ Sanchez-Yamagishi, K.\ Watanabe, T.\ Taniguchi, E.\ Kaxiras,
R.\ C.\ Ashoori, P.\ Jarillo-Herrero, Nature {\bf 556}, 80  (2018).

\bibitem{jh3}
S.\ Carr, S.\ Fang, P.\ Jarillo-Herrero, E.\ Kaxiras, Phys.\ Rev.\ B
{\bf 98}, 085144 (2018).

\bibitem{yankovitz} M.\ Yankowitz, S.\ Chen, H.\ Polshyn, K.\ Watanabe,
T.\ Taniguchi, D.\ Graf, A.\ F.\ Young, C.\ R.\ Dean,
arxiv:1808.07865

\bibitem{arnold}
F.\ Arnold, J.\ Ny\'eki, J.\ Saunders, arXiv:1804.00179.

\bibitem{volovik1}
G.\ E.\ Volovik, JETP Letters 107, 516 (2018), arXiv:1803.08799.

\bibitem{volovik2}
N.\ Kopnin, T.\ Heikkil\"a, G.\ E.\ Volovik, Phys.\ Rev.\ B {\bf 83}, 220503(R)
(2011).

\bibitem{PLA2018}V.\ A.\ Khodel, J.\ W.\ Clark, M.\ V.\ Zverev,
Phys. Lett. A{\bf 382}, 3281 (2018).

\bibitem{cooper}
R.\ A.\ Cooper, Y.\ Wang, B.\ Vignolle, O.\ J.\ Lipscombe, S.\ M.\ Hayden,
Y.\ Tanabe, T.\ Adachi, Y.\ Koike, M.\ Nohara, H.\ Takagi, C.\ Proust,
N.\ E.\ Hussey, Science {\bf 323}, 603 (2009).

\bibitem{taillefer}
L.\ Taillefer, Annu.\ Rev.\ Condens.\ Matter Phys.\ {\bf 1}, 51 (2010).

\end{thebibliography}
\end{document}